\documentclass[10pt,twocolumn,letterpaper]{article}

\usepackage[pagenumbers]{cvpr} 

\usepackage{algorithm}
\usepackage{algorithmic}

\usepackage{tabularray}
\usepackage{color}
\usepackage{graphicx}
\usepackage{adjustbox}
\usepackage{amsmath,amssymb,amsfonts}

\definecolor{cvprblue}{rgb}{0.21,0.49,0.74}
\usepackage[pagebackref,breaklinks,colorlinks,allcolors=cvprblue]{hyperref}

\def\paperID{15372} 
\def\confName{CVPR}
\def\confYear{2026}

\title{DLADiff: A Dual-Layer Defense Framework against Fine-Tuning and Zero-Shot Customization of Diffusion Models}

\author{
Jun Jia$^1$\quad
Hongyi Miao$^2$\quad 
Yingjie Zhou$^1$\quad
Linhan Cao$^1$\quad
Yanwei Jiang$^1$\quad
Wangqiu Zhou$^3$\quad\\
Dandan Zhu$^4$\quad
Hua Yang$^1$\quad
Wei Sun$^4$\quad
Xiongkuo Min$^1$\quad
Guangtao Zhai$^1$\\
$^1$Shanghai Jiao Tong University\quad$^2$Shandong University\\
$^3$Hefei University of Technology\quad$^4$East China Normal University\\
{\tt\small jiajun0302@sjtu.edu.cn}} 

\begin{document}
\maketitle
\begin{abstract}
With the rapid advancement of diffusion models, a variety of fine-tuning methods have been developed, enabling high-fidelity image generation with high similarity to the target content using only 3 to 5 training images. More recently, zero-shot generation methods have emerged, capable of producing highly realistic outputs from a single reference image without altering model weights. However, technological advancements have also introduced significant risks to facial privacy. Malicious actors can exploit diffusion model customization with just a few or even one image of a person to create synthetic identities nearly identical to the original identity. Although research has begun to focus on defending against diffusion model customization, most existing defense methods target fine-tuning approaches and neglect zero-shot generation defenses. To address this issue, this paper proposes \textbf{D}ual-\textbf{L}ayer \textbf{A}nti-\textbf{Diff}usion (DLADiff) to defense both fine-tuning methods and zero-shot methods. DLADiff contains a dual-layer protective mechanism. The first layer provides effective protection against unauthorized fine-tuning by leveraging the proposed Dual-Surrogate Models (DSUR) mechanism and Alternating Dynamic Fine-Tuning (ADFT), which integrates adversarial training with the prior knowledge derived from pre-fine-tuned models. The second layer, though simple in design, demonstrates strong effectiveness in preventing image generation through zero-shot methods. Extensive experimental results demonstrate that our method significantly outperforms existing approaches in defending against fine-tuning of diffusion models and achieves unprecedented performance in protecting against zero-shot generation.

\end{abstract}    
\section{Introduction}
\label{sec:introduction}

\begin{figure}[htbp]
\centering
\includegraphics[width=0.48\textwidth]{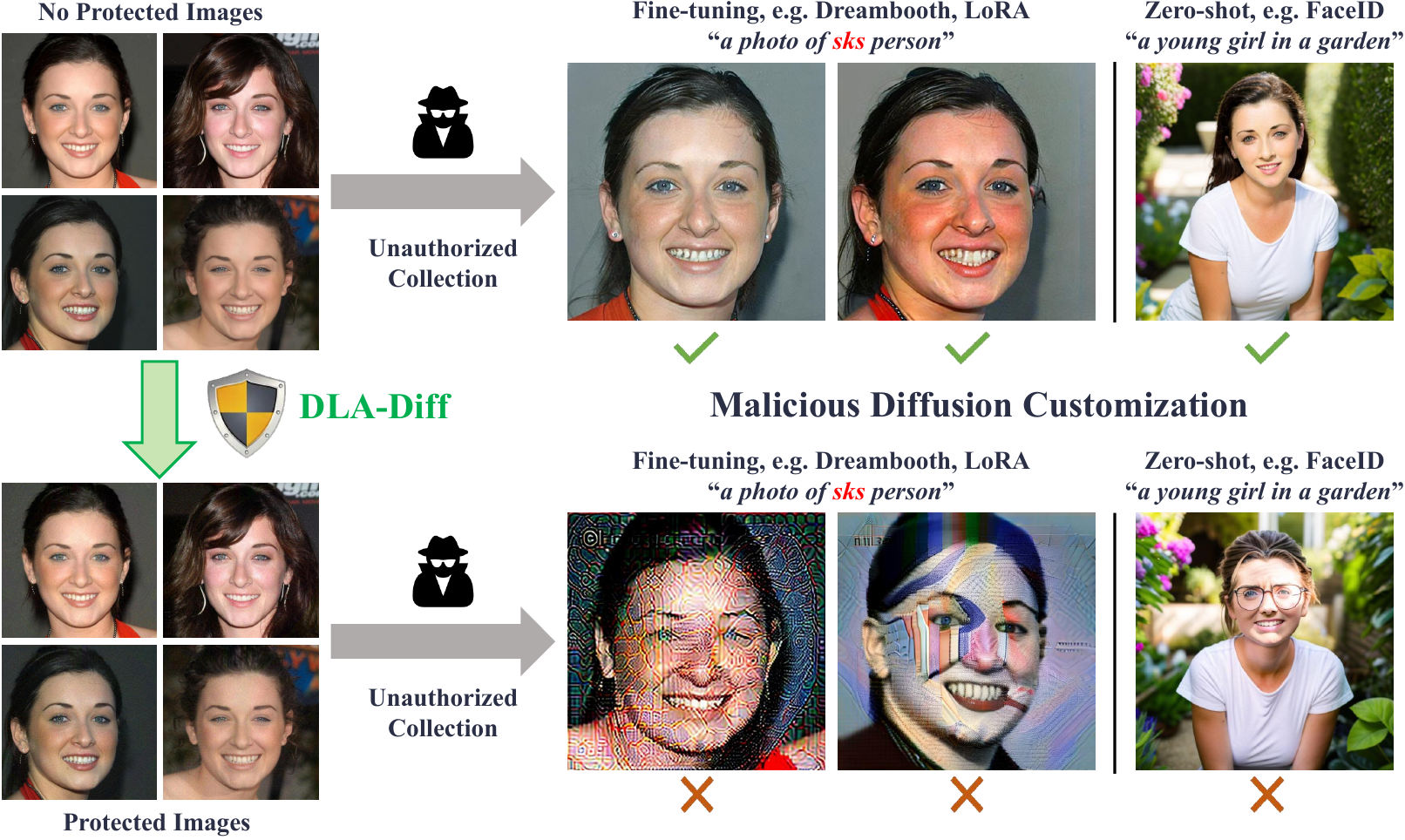}
\caption{The DLADiff framework protects personal photos by simultaneously resisting fine-tuning and zero-shot generation in diffusion models, significantly degrading the output quality of maliciously customized models. \textbf{\textcolor{red}{Some of the visualization results in this paper may cause discomfort to viewers.}}}
\label{fig:application}
\end{figure}

In recent years, diffusion models have emerged as the dominant paradigm in image generation, consistently demonstrating superior performance in producing high-fidelity visual content across a wide range of applications. To facilitate the adaptation of diffusion models to specific and data-limited settings, a variety of fine-tuning methods have been developed. These methods, such as DreamBooth~\cite{ruiz2023dreambooth} and LoRA~\cite{hu2022lora}, leverage small-scale datasets to effectively fine-tune pretrained model weights, thereby enabling the high-fidelity generation of specific attributes, including facial identities and image styles. The dataset used for fine-tuning typically comprises a limited number of images, with some cases involving as few as three to five samples. More recently, zero-shot methods based on diffusion models have been proposed. These methods enable the generation of content that is highly consistent with the target, relying solely on a single reference instance. The development of these methods have progressively reduced the diffusion model's reliance on large-scale prior data, thereby improving its applicability in real-world scenarios. 

However, technological advancements also significantly reduce the cost of producing deepfake contents, thereby giving rise to non-negligible ethical and privacy risks. Unauthorized individuals, such as hackers, can generate fake facial identities using just three to five or even one private photo through diffusion model fine-tuning and zero-shot generation methods. To mitigate this security vulnerability, a series of defensive approaches termed anti-diffusion model customization (denoted as anti-customization) are proposed. By introducing protective perturbations to original portraits, these methods degrade the quality of outputs generated through diffusion model fine-tuning. However, as zero-shot generation methods continue to advance, the limitations of existing anti-customization approaches have become increasingly apparent, particularly their inability to effectively generalize to zero-shot generation scenarios. To defend against diverse customizations of diffusion models in practical scenarios, this paper presents, for the first time, a systematic defense framework capable of simultaneously prevent identity theft through both fine-tuning and zero-shot generation methods, which is illustrated in Figure~\ref{fig:application}.

We first systematically analyze the differences between fine-tuning methods and zero-shot methods. Diffusion model fine-tuning dynamically updates the pre-trained weights, whereas zero-shot methods incorporate a pretrained encoder to extract identity embeddings and inject them into the diffusion model's noise predictor via additional cross-attention layers. Therefore, the essence of defense against fine-tuning lies in misleading the fine-tuning process into capturing erroneous patterns, thereby rendering genuine identity information and facial structures unlearnable. In contrast to that, the principle of defending against zero-shot methods is more closely aligned with generating adversarial samples targeting a fixed pretrained identity encoder, while requiring consideration of generalization across diverse encoders. Furthermore, given that the identity encoder employed in zero-shot methods is decoupled from the model weights subject to fine-tuning, it is natural to propose employing two distinct layers of protective perturbations to defend against fine-tuning and zero-shot methods separately. 

Based on the aforementioned analysis, this paper proposes \textbf{D}ual-\textbf{L}ayer \textbf{A}nti-\textbf{Diff}usion (DLADiff) to defense both fine-tuning methods and zero-shot methods, which contains a dual-layer protective mechanism. The first layer of protective perturbations is designed to defend against fine-tuning methods. Building upon existing anti-customization methods targeting fine-tuning methods, we propose, for the first time, a dual-surrogate-based alternating dynamic optimization framework that significantly enhances the protection of facial local details through pre-fine-tuning a static surrogate model. The second layer of protective perturbations targets zero-shot methods and employs a weighted Projected Gradient Descent (PGD) attack to ensure generalization across diverse zero-shot methods. The main contributions of this work include:
\begin{itemize}
\item We propose DLADiff, the first dual-layer anti-customization framework against both diffusion model fine-tuning methods and zero-shot methods.
\item We significantly improve the defense performance against fine-tuning methods by introducing Dual-Surrogate Models (DSUR) mechanism and Alternating Dynamic Fine-tuning (ADFT).
\item We propose a simple yet effective defense mechanism to enhance the generalization capability of diverse zero-shot generation methods.
\end{itemize}


\section{Related Work}
\label{sec:related_work}

\subsection{Customization of Diffusion Models}
To enable pre-trained diffusion model weights to generate user-specified images, the technique of diffusion model fine-tuning has been introduced~\cite{ruiz2023dreambooth,kumari2023multi}. Among these typical fine-tuning methods, DreamBooth~\cite{ruiz2023dreambooth} optimizes the weights of the UNet and text encoder, LoRA~\cite{hu2022lora} introduces fine-tuned low-rank matrices into the pretrained weights, and Textual Inversion~\cite{gal2022image} learns to optimize adaptive text embeddings. However, fine-tuning methods rely on multiple images depicting a specific subject or style, which are often difficult to obtain in real-world scenarios. To overcome this limitation, zero-shot image-to-image generation has been introduced, which rely solely on a single reference image to generate visually consistent content. These methods employ an image encoder to derive embeddings from the reference image and utilize cross-attention layers to incorporate these features into designated layers of the UNet. For general generation, IP-Adapter~\cite{ye2023ip-adapter} employs CLIP~\cite{radford2021learning} as the image encoder. For facial identity generation, IP-Adapter Faceid~\cite{ye2023ip-adapter} and Instant-ID~\cite{wang2024instantid} encode embeddings through pretrained ArcFace~\cite{deng2019arcface} models. Recent methods such as Photomakerr~\cite{li2024photomaker}, PULID~\cite{guo2024pulid}, and StoryMaker~\cite{zhou2024storymaker} further integrate both CLIP and ArcFace encoders to enhance identity preservation. Compared to fine-tuning, zero-shot approaches reduce the reliance on multiple training images, thus improving their feasibility in real-world applications.

\subsubsection{Defense Methods for Customization}

The widespread use of diffusion model customization has raised concerns about unauthorized misuse of personal images. To address this risk, numerous anti-customization methods are proposed to protect copyrighted content, such as artistic styles~\cite{shan2023glaze,shan2024nightshade} and facial identities~\cite{van2023anti}, from being reproduced. Since zero-shot methods are newly emerging, most existing anti-customization methods focus on defending against fine-tuning. MIST~\cite{liang2023adversarial} targets diffusion model fine-tuning by adding pixel-level adversarial noise to original images, causing the model to generate a predefined noisy output. CAAT~\cite{xu2024perturbing} shows that small perturbations in the attention mechanism can strongly misdirect fine-tuning. ACE~\cite{zheng2023understanding} introduces a unified target to guide perturbation optimization in both forward encoding and reverse generation, effectively reducing offset issues and enhancing protection robustness and transferability. Anti-DreamBooth~\cite{van2023anti} introduces a dynamically updated surrogate model to enhance robustness. This work inspires a variety of subsequent methods based on adversarial training~\cite{zheng2025anti,liu2024disrupting,sun2025pretender}. Pretender~\cite{sun2025pretender} further proves that the introduction of adversarial training framework can effectively fools downstream fine-tuning tasks and works across diverse fine-tuning methods.
\section{Preliminaries}
\label{sec:preliminaries}

\subsection{Background}
\noindent\textbf{Diffusion Models} are currently the most widely used image generative model, with a training process composed of two decoupled phases: the forward and backward processes. Given an input image $x_0$, the forward process progressively adds standard Gaussian noise to $x_0$ at each timestep $t$ through a Markov chain. The output $x_t$ at each timestep is defined as follows:
\begin{equation}
\begin{aligned}
&x_t=\sqrt{\overline{\alpha}_t}x_0+\sqrt{1-\overline{\alpha}_t}\epsilon, \\
\end{aligned}
\label{eq:1}
\end{equation}
where $\alpha_t=1-\beta_t$, $\overline{\alpha}_t=\prod \limits_{s=1}^t\alpha_s$, and $\epsilon\sim\mathcal{N}(0,\mathbf{I})$. After $T$ steps, the clean input $x_0$ is transformed into pure Gaussian noise. In contrast to the forward process, the backward process employs a learnable neural network $\epsilon_{\theta}(x_{t+1},t)$ to estimate the noise added at the current time step from $x_{t+1}$ and thereby reconstruct the variables at the previous time step, $x_{t}$, through denoising. The network weights $\theta$ of $\epsilon_{\theta}$ are optimized by minimizing the following loss function:
\begin{equation}
\begin{aligned}
&\mathcal{L}_{ucond}(\theta,x_0)=\mathbb{E}_{x_0,t,\epsilon\sim\mathcal{N}(0,\mathbf{I})}||\epsilon-\epsilon_\theta(x_{t+1},t)||^2_2, \\
\end{aligned}
\label{eq:2}
\end{equation}
where $\epsilon$ is the reference noise added in the forward process. 

As an extension of diffusion models, stable diffusion models perform the noise addition and denoising processes in the latent space $\mathcal{Z}$ of a pretrained variational autoencoder~\cite{vae} ($\mathbf{VAE}$), thereby reducing computational costs. By incorporating text prompts as conditional inputs, it enables effective text-guided image generation. The objective of stable diffusion models is formulated as follows:
\begin{equation}
\begin{aligned}
&\mathcal{L}_{cond}(\theta,z_0)=\mathbb{E}_{z_0,t,c,\epsilon\sim\mathcal{N}(0,\mathbf{I})}||\epsilon-\epsilon_\theta(z_{t+1},t,c)||^2_2, \\
\end{aligned}
\label{eq:3}
\end{equation}
where $c$ represents the text prompt condition and $z_0$ denotes the latent variables of the input images.

\noindent\textbf{Diffusion Fine-tuning Methods} involve optimizing all or part of the model weights using a small-scale dataset, enable the generation of content that closely resembles the images used during fine-tuning. DreamBooth is a widely adopted fine-tuning method for stable diffusion models. Given a set of images sharing common characteristics, such as the same facial identity, and a text prompt $c$ containing a specific trigger word, Dreambooth enforces a strong association between the fine-tuning images and the trigger word, enabling the model to generate images of that specific identity in response to the trigger word during inference. In addition, to mitigate model overfitting, DreamBooth adds a regularization term during fine-tuning that uses prior prompts $c'$, text inputs without the trigger word, and corresponding images from the original weights. We define $c$ and $c'$ as ``\textit{a photo of sks person}" and ``\textit{a photo of person}", repectively. The objective of DreamBooth is formulated as follows:
\begin{equation}
\begin{aligned}
\mathcal{L}_{db}(\theta,z_0)=&\mathbb{E}_{z_0,t,t'}||\epsilon-\epsilon_\theta(z_{t+1},t,c)||^2_2 \\
&+\lambda||\epsilon'-\epsilon_\theta(z'_{t+1},t',c')||^2_2. \\
\end{aligned}
\label{eq:4}
\end{equation}
where $\epsilon,\epsilon'\sim\mathcal{N}(0,\mathbf{I})$, $t,t'\in[1,T]$, and $\lambda$ balances the weights of regularization term. 

Low-Rank Adaptation (LoRA) preserves the original model weights by freezing them during training, while selectively fine-tuning low-rank matrices injected into the attention layers. This approach significantly reduces computational overhead and alleviates overfitting risks compared to full-parameter fine-tuning.

\noindent\textbf{Zero-shot Image-to-Image Generation} is recently proposed to capture specific image features from a single reference image without altering the pretrained model weights. These methods employ an image encoder to extract embeddings from the reference image and utilize additional cross-attention layers to incorporate them into designated layers of the UNet architecture.

\begin{figure*}[htbp]
\centering
\includegraphics[width=\textwidth]{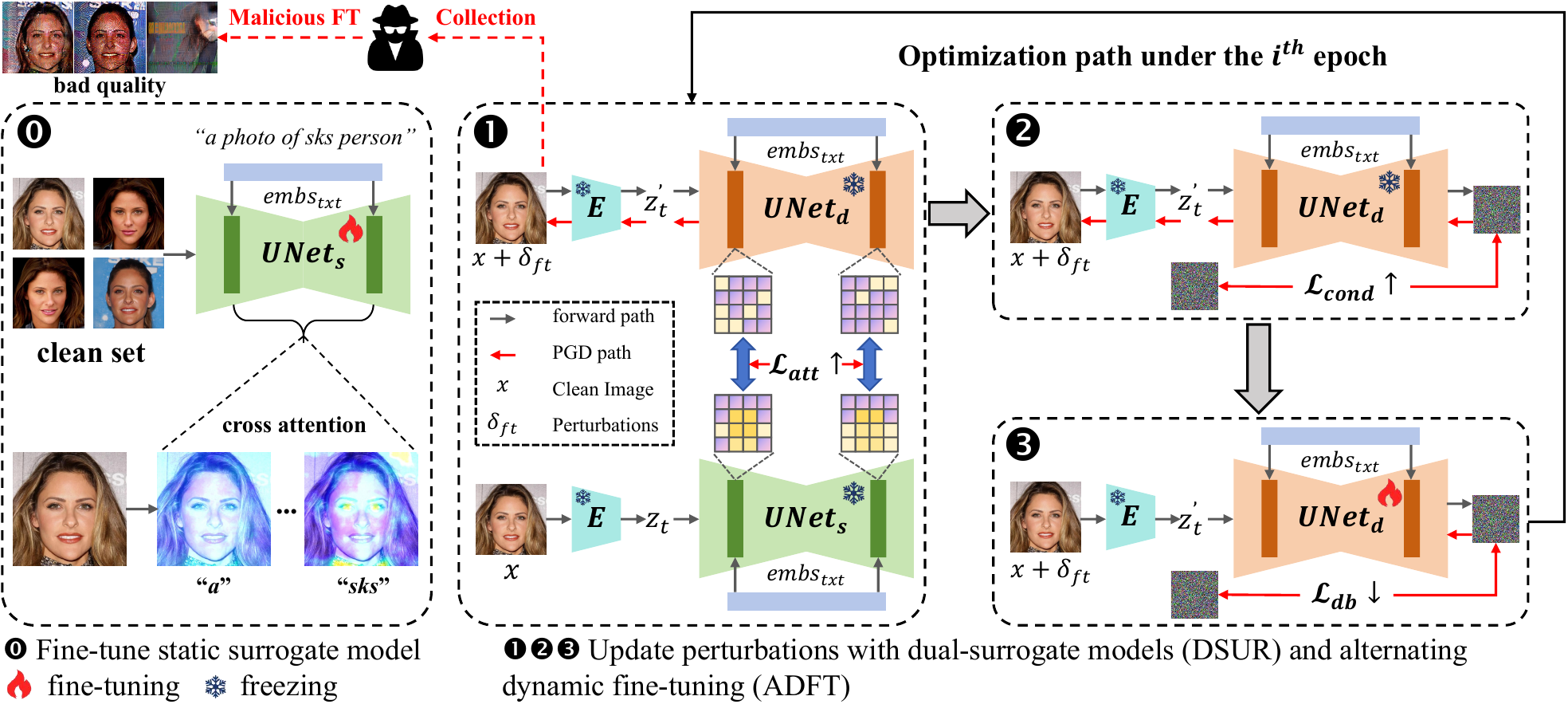}
\caption{The optimization process of the first layer of protective perturbation in DLADiff. This layer can effectively defense fine-tuning based diffusion model customization. The optimization process includes four steps. \textbf{Step-0} involves the pre-fine-tuning of a static surrogate model, denoted as $\mathbf{UNet_s}$, using a clean dataset that shares the same identity as the images to be protected. \textbf{Step-1} optimizes perturbations $\delta_{ft}$ by disrupting the attention maps using $\mathbf{UNet_s}$ as reference. \textbf{Step-2} and \textbf{Step-3} involve the optimization based on adversarial training. Repeat \textbf{Step-1} to \textbf{Step-3} until the preset epoch is reached.}
\label{fig:pipeline-ft}
\end{figure*}

\subsection{Problem Definition\label{problem_definition}}
Let $\mathcal{X}=\{x_1,x_2,...,x_n\}$ denote a set of personal portrait images requiring protection. Our method aims to generate a protective perturbation $\delta_i$ and generate the perturbed version $x'_i=x_i+\delta_i$ for each image in $\mathcal{X}$, such that the perturbed dataset $\mathcal{X}'$ can be safely published. When unauthorized users access $\mathcal{X}'$ and attempt to use these images for diffusion model customization, the resulting outputs exhibit severely degraded quality, effectively preventing identity theft. We define the dataset perturbation as $\delta^*$. The aforementioned objective can be formulated as follows:
\begin{equation}
\begin{aligned}
&\delta^*=\mathop{\arg\min}\limits_{\delta^*}\ \mathcal{A}(\mathbf{DM_c},\mathcal{X}'),\\
&s.t.\ \ ||\delta^*||_\infty\leq\eta, 
\end{aligned}
\label{eq:5}
\end{equation}
where $\mathbf{DM_c}$ represents the customized diffusion models, $\mathcal{A}$ denotes a metric for evaluating generation quality, e.g. Fréchet Inception Distance~\cite{heusel2017gans} (FID) and Identity Score Matching~\cite{deng2019arcface} (ISM), and $\eta$ is the bound of perturbation. In practice, unauthorized users may employ both fine-tuning and zero-shot methods to customize generated results. Therefore, the perturbation $\delta^*$ must provide defense across both fine-tuning and zero-shot methods. This problem is inherently more challenging than the scenario limited to a single type of diffusion model customization. In the case of fine-tuning methods such as DreamBooth, unauthorized users would apply the loss function defined in Eq.~\ref{eq:4} to fine-tune on $\mathcal{X}'$, under which condition $\mathbf{DM_c}$ can be formulated as $\epsilon_{\theta^*}$. $\theta^*$ is the weights after fine-tuning:
\begin{equation}
\begin{aligned}
\theta^*= \mathop{\arg\min}\limits_{\theta}\sum\limits_{x\in\mathcal{X'}}\mathcal{L}_{db}(\theta,x).
\end{aligned}
\label{eq:6}
\end{equation}
For zero-shot methods, an unknown identity encoder is employed to extract identity embeddings from $\mathcal{X}'$, and $\mathbf{DM_c}$ can accordingly be represented as $\mathbf{IE}$. For convenience and compatibility, pretrained face recognition models such as ArcFace are commonly used as identity encoders.

\section{Methodology}
\label{sec:methodology}
To address the challenges mentioned in Section~\ref{problem_definition}, this section presents the proposed defense approach, named \textbf{D}ual-\textbf{L}ayer \textbf{A}nti-\textbf{D}iffusion (DLADiff), in detail. Based on the preceding analysis, fine-tuning of diffusion models involves dynamic adjustment of model weights, whereas zero-shot diffusion generation methods typically employ a pretrained face recognition model as the identity encoder. Consequently, defending against fine-tuning demands more sophisticated and carefully optimized perturbations. Leveraging this insight, our approach incorporates a dual-layer perturbation defense mechanism: the first layer is specifically designed to counter fine-tuning-based customization, while the second layer targets zero-shot methods. Reversing the order may erase the perturbations for zero-shot methods when optimizing the other layer of perturbations, weakening the defense against zero-shot generation.

\subsection{Perturbations for Fine-tuning Methods}
The prevailing mainstream approaches employ an adversarial training-based framework~\cite{van2023anti}. The primary objective of these methods is to optimize perturbations via Projected Gradient Descent~\cite{madry2018towards} (PGD) attacks, such that the noise predicted by the UNet~\cite{ronneberger2015u} diverges from the noise introduced during the forward diffusion process. This process effectively maximize $\mathcal{L}_{cond}$ in Eq.~\ref{eq:3}. To simulate the fine-tuning procedure, these approaches incorporate a dynamically updated surrogate model. The optimization of the surrogate model and the protective perturbation is performed in an alternating manner, thereby improving the robustness of perturbations. The perturbations introduced by these methods may lead to overfitting during fine-tuning, mislead the optimization trajectory, and consequently result in blocky artifacts in the generated images. However, these methods inadequately disrupt facial details, prompting the development of attention map disruption-based approaches. These approaches mislead fine-tuning process by disrupting UNet's self-attention and cross-attention maps. However, their performance improvement is limited in practice. This limitation stems from the use of attention features derived from a surrogate model initialized with pretrained weights, which deviates significantly from the ideal attention pattern observed after fine-tuning.

Building upon the analysis of limitations in existing anti-fine-tuning methods, the first layer of protective perturbations $\delta_{ft}$ employs an optimization strategy grounded in a dual-surrogate model framework, as illustrated in Figure~\ref{fig:pipeline-ft}. This framework consists of two stages: \textbf{(1)} Fine-tuning the static surrogate model, \textbf{(2)} Updating perturbations with Dual-Surrogate Models (DSUR) and Alternating Dynamic Fine-Tuning (ADFT).

\subsubsection{Dual-Surrogate Models (DSUR) Mechanism}
This mechanism combines a dynamically updated surrogate model with a fully fine-tuned surrogate model with fixed weights, improving the perturbations' effectiveness in disrupting both global and local facial features. As illustrated in Figure~\ref{fig:pipeline-ft}, we first fine-tune a static surrogate model on a clean dataset $\mathcal{X}_{clean}$, which contains multiple portraits sharing the same identity as the images to be protected. The clean dataset can include the images to be protected. To ensure broad applicability, the static surrogate model is trained using DreamBooth, with the text prompt ``\textit{a photo of sks person}". To reduce computational overhead, only the UNet weights are updated, yielding the fine-tuned $\mathbf{UNet_s}$. The weights $\theta_s$ of $\mathbf{UNet_s}$ are optimized as follows:
\begin{equation}
\begin{aligned}
\theta_s=&\mathop{\arg\min}\limits_{\theta}\ \mathbb{E}_{z_0,t,c}||\epsilon-\mathbf{UNet_s}(z_{t+1},t,c,\theta)||^2_2, \\
&s.t.\ z_0=\mathbf{Enc_{vae}}(x)\ and\ x\in \mathcal{X}_{clean},
\end{aligned}
\label{eq:7}
\end{equation}
where $\mathbf{Enc_{vae}}$ denotes the VAE encoder, $z_0$ represents the latent variables of the images, and the definitions of all other terms are consistent with those in Eq.~\ref{eq:4}. We omit the prior regularization term of Eq.~\ref{eq:4} for simplicity. After fine-tuning, the cross-attention layers in $\mathbf{UNet_s}$ are able to selectively attend to the key tokens in the text prompt, specifically ``\textit{sks}". As illustrated in Figure~\ref{fig:pipeline-ft}, the cross-attention map associated with ``\textit{sks}" exhibits high activation values in the eye, nose, and mouth regions of the portrait. There is a strong correlation between these regions and the facial identity. Furthermore, the self-attention layers in $\mathbf{UNet_s}$ also concentrate on capturing the structural features of the portrait. Then, we define a attention loss function as $\mathcal{L}_{att}$ based on both the static and dynamic surrogate models:
\begin{equation}
\begin{aligned}
\mathcal{L}_{att}(\theta_s,\theta_d,x,\delta_{ft})=&||\mathcal{M}c_{\theta_s}(x)-\mathcal{M}c_{\theta_d}(x+\delta_{ft})||^2_2\\
+||&\mathcal{M}s_{\theta_s}(x)-\mathcal{M}s_{\theta_d}(x+\delta_{ft})||^2_2,
\end{aligned}
\label{eq:8}
\end{equation}

where $\mathcal{M}c_{\theta_s}(x)$ and $\mathcal{M}s_{\theta_s}(x)$ denote the cross-attention and self-attention maps, respectively, of the clean image $x$ associated with the static surrogate model $\mathbf{UNet_s}$. Analogously, $\mathcal{M}c_{\theta_d}(x+\delta_{ft})$ and $\mathcal{M}s_{\theta_d}(x+\delta_{ft})$ are the attention maps of the perturbed image $x+\delta_{ft}$ from the dynamic surrogate model $\mathbf{UNet_d}$. $\mathbf{UNet_d}$ is initialized from a pretrained weights without any instance-related knowledge.

\subsubsection{Alternating Dynamic Fine-Tuning (ADFT)}
To fully exploit the strengths of the two surrogate models, we update the perturbations $\delta_{ft}$ in two sequential stages and alternately optimize the dynamic surrogate model. 

\noindent\textbf{Stage-1:} In the first stage, we fix the weights of both $\mathbf{UNet_s}$ and $\mathbf{UNet_d}$, and optimize $\delta_{ft}$ along the direction of the gradient ascent of $\mathcal{L}_{att}$ through Projected Gradient Descent (PGD) attacks. This operation increases the value of $\mathcal{L}_{att}$ to introduce resistance to the model parameter updates when approaching the ideal attention state. 

\noindent\textbf{Stage-2:} In the second stage, we optimize $\delta_{ft}$ along the direction of the gradient ascent of $\mathcal{L}_{cond}$. This operation only uses the dynamic surrogate model $\mathbf{UNet_d}$. 

Finally, we optimize the dynamic surrogate model $\mathbf{UNet_d}$ by minimizing $\mathcal{L}_{db}$ on perturbed images $x+\delta_{ft}$. The aforementioned three stages constitute one epoch of perturbation optimization. By iteratively repeating multiple epochs, the final perturbations is obtained. One epoch of ADFT can be concisely formulated as follows:

\begin{equation}
\begin{aligned}
\delta_{ft}&\leftarrow\mathop{\arg\max}\limits_{\delta_{ft}}\ \mathcal{L}_{att}(\theta_s,\theta_d,x,\delta_{ft}),\\
\delta_{ft}&\leftarrow\mathop{\arg\max}\limits_{\delta_{ft}}\ \mathcal{L}_{cond}(\theta_d,x+\delta_{ft}),\\
\theta_d&\leftarrow\mathop{\arg\min}\limits_{\theta}\ \mathcal{L}_{db}(\theta_d,x+\delta_{ft}),\\
&s.t.\ ||\delta_{ft}||_\infty\leq\eta_{ft},
\end{aligned}
\label{eq:9}
\end{equation}
where $\eta_{ft}$ is the bound of $\delta_{ft}$. The complete optimization algorithm is provided in supplementary materials.  

\begin{table*}
\caption{Comparison results with state-of-the-art methods against Dreambooth fine-tuning. We evaluate these methods on two inference prompts. The best and second-best results are marked by \textcolor{red}{red} and \textcolor{blue}{blue}.}
\centering
\begin{adjustbox}{width=\textwidth}
\begin{tblr}{
  row{1} = {c},
  row{2} = {c},
  cell{1}{1} = {r=2}{},
  cell{1}{2} = {r=2}{},
  cell{1}{3} = {c=5}{},
  cell{1}{8} = {c=5}{},
  cell{3}{1} = {r=6}{c},
  cell{3}{2} = {c},
  cell{4}{2} = {c},
  cell{5}{2} = {c},
  cell{6}{2} = {c},
  cell{7}{2} = {c},
  cell{8}{2} = {c},
  cell{9}{1} = {r=6}{c},
  cell{9}{2} = {c},
  cell{10}{2} = {c},
  cell{11}{2} = {c},
  cell{12}{2} = {c},
  cell{13}{2} = {c},
  cell{14}{2} = {c},
  vline{3,8} = {1-15}{},
  hline{1,15} = {-}{0.08em},
  hline{3,9} = {-}{},
  hline{4} = {2-12}{},
  hline{10} = {2-12}{},
}
Dataset   & Method         & ``\textit{a photo of sks person}''  &     &     &         &    & ``\textit{a dslr portrait of sks person}''  &     &     &         &    \\
          &                & FDR$\downarrow$ & ISM$\downarrow$ & FID$\uparrow$ & FIQA$\downarrow$ & MOS$\downarrow$ & FDR$\downarrow$ & ISM$\downarrow$ & FID$\uparrow$ & FIQA$\downarrow$ & MOS$\downarrow$ \\
CelebA-HQ & w/o Protect    & 0.996 & 0.580 & 53.44 & 0.385 & N/A & 0.934 & 0.364 & 92.88 & 0.436 & N/A \\
          & MIST           & 0.980 & 0.516 & 94.49 & 0.252 & 3.36 & 0.948 & 0.368 & 106.1 & 0.309 & 3.43   \\
          & Anti-DB        & 0.851 & 0.452 & 144.4 & 0.235 & 2.63 & 0.892 & 0.328 & \textcolor{blue}{166.4} & 0.280 & 2.41   \\
          & DisDiff        & \textcolor{blue}{0.482} & \textcolor{blue}{0.241} & \textcolor{blue}{201.8} & \textcolor{red}{0.207} & \textcolor{blue}{1.79} & \textcolor{blue}{0.861} & \textcolor{blue}{0.322} & 145.2 & \textcolor{red}{0.242} & \textcolor{blue}{2.32}   \\
          & Anti-diffusion & 0.802 & 0.425 & 164.6 & 0.239 & 2.37 & 0.906 & 0.350 & 138.8 & 0.281 & 2.56   \\
          & Ours           & \textcolor{red}{0.201} & \textcolor{red}{0.096} & \textcolor{red}{233.7} & \textcolor{blue}{0.225} &  \textcolor{red}{1.57} & \textcolor{red}{0.668} & \textcolor{red}{0.264} & \textcolor{red}{187.9} & \textcolor{blue}{0.252} & \textcolor{red}{1.68}   \\
VGGFace2  & w/o Protect    & 0.928 & 0.521 & 62.53 & 0.383 & N/A & 0.907 & 0.397 & 93.70 & 0.423 & N/A  \\
          & MIST           & 0.844 & \textcolor{blue}{0.268} & 175.8 & 0.270 & 2.55 & 0.822 & \textcolor{red}{0.257} & \textcolor{blue}{186.9} & 0.273 & 2.12   \\
          & Anti-DB        & \textcolor{blue}{0.677} & 0.300 & 186.7 & 0.220 & 1.98 & \textcolor{red}{0.746} & \textcolor{blue}{0.265} & \textcolor{red}{200.7}    & \textcolor{red}{0.217} & \textcolor{red}{1.86}   \\
          & DisDiff        & 0.741 & 0.362 & \textcolor{blue}{187.4} & \textcolor{red}{0.201} & \textcolor{blue}{1.83} & 0.880 & 0.375 & 137.7 & 0.240 & 2.43   \\
          & Anti-diffusion & 0.824 & 0.318 & 165.1 & 0.238 & 2.46 & 0.842 & 0.329 & 160.7 & 0.243 & 2.54   \\
          & Ours           & \textcolor{red}{0.608} & \textcolor{red}{0.263} & \textcolor{red}{194.0} & \textcolor{blue}{0.217} & \textcolor{red}{1.76} & \textcolor{blue}{0.807} & 0.310 & 177.1 & \textcolor{blue}{0.221} & \textcolor{blue}{2.09}   \\   
\end{tblr}
\end{adjustbox}
\label{tab:compare}
\end{table*}

\subsection{Perturbations for Zero-shot Methods}
Compared to fine-tuning methods, diffusion-based zero-shot generation methods utilize pretrained identity encoders to extract embeddings, which are then injected into the UNet architecture via additional cross-attention layers. Since the identity encoder weights are fixed, defending against zero-shot methods is more like creating adversarial samples than unlearnable samples. Based on this sight, we design simple yet effective second-layer protective perturbations to defend zero-shot methods. 

Given an image $x'$ protected by the first layer, we first extract the facial region as $x'_f$ via face alignment, a necessary preprocessing step for identity encoding. The same procedure is applied to the unprotected image $x$ to obtain $x_f$. We denote the perturbations targeting zero-shot methods as $\delta_{zs}$ and define a loss function to evaluate the identity similarity between the perturbed and original facial identities:
\begin{equation}
\begin{aligned}
\mathcal{L}_{id}=1-\sum^{N}_{i=1} \mathbf{CosSim}(\mathbf{IE}_{i}(x'_f+\delta_{zs}), \mathbf{IE}_{i}(x_f)),
\end{aligned}
\label{eq:10}
\end{equation}
where $\mathbf{IE}_{i}$ denotes the $i^{th}$ identity encoder employed in the optimization process. We select $N$ distinct encoders and weight their corresponding similarity scores to enhance the generalization capability of $\delta_{zs}$. Then, the perturbations $\delta_{zs}$ are updated by the gradients $\nabla$ through PGD attacks:
\begin{equation}
\begin{aligned}
&\delta_{zs}=\delta_{zs}+\sigma_{zs}*\nabla_{\mathcal{\delta}_{zs}}\mathcal{L}_{id},
\ s.t.\ ||\delta_{zs}||_\infty\leq\eta_{zs}, 
\end{aligned}
\end{equation}
where $\sigma_{zs}$ is the optimization stride and $\eta_{zs}$ is the bound of $\delta_{zs}$. Finally, $x'_f+\delta_{zs}$ is transformed into the original coordinate system through the inverse transformation used in face alignment. In practice, the introduction of perturbations may cause a minor influence on the landmark coordinates. To enhance robustness to landmark detection, we introduce slight random noise into the affine matrix used in face alignment. The detailed pipeline of this process is provided in supplementary materials.
\section{Experiments}
\label{sec:experiments}
\subsection{Experimental Settings}

\begin{figure*}[htbp]
\centering
\includegraphics[width=\textwidth]{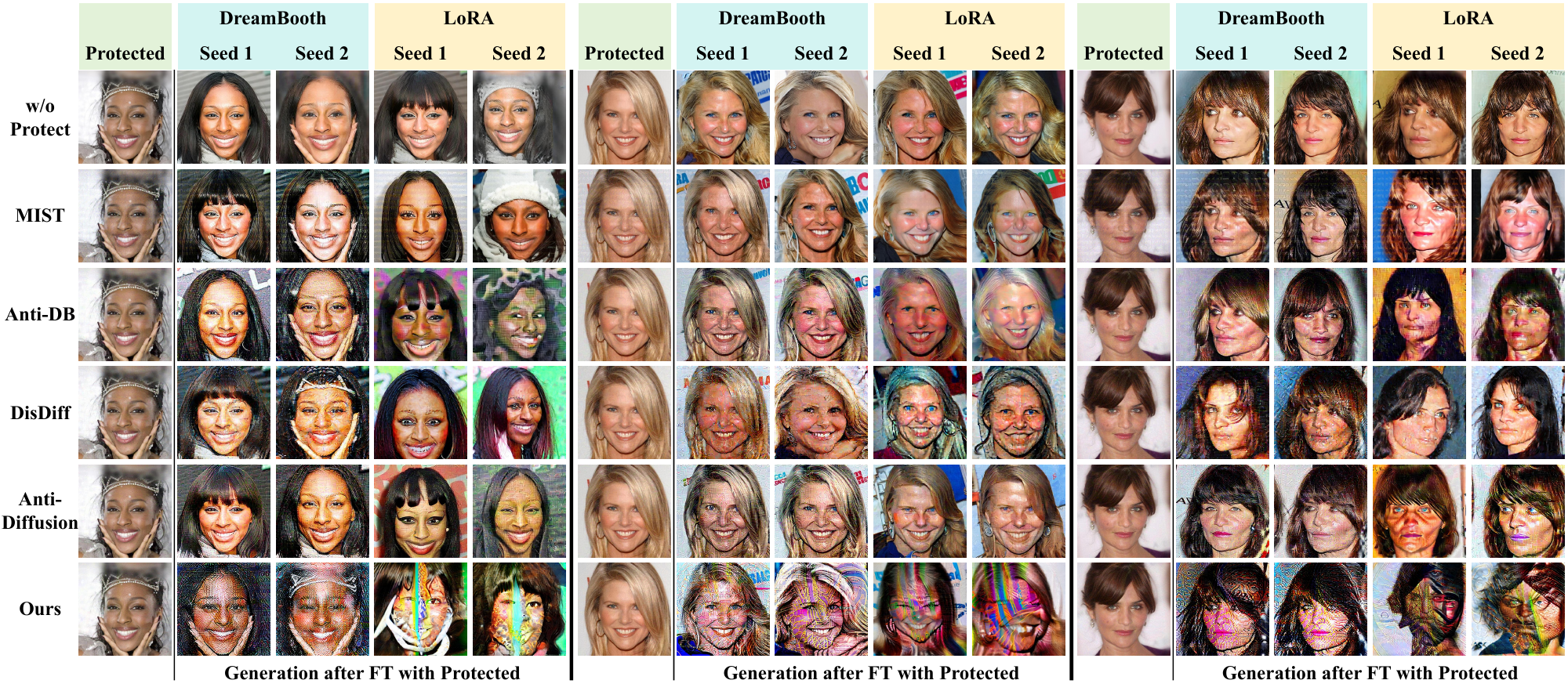}
\caption{The comparison results on defending DreamBooth and LoRA Fine-tuning. We use the protected images to fine-tune a pretrained stale diffusion model. Then, we generate images under diverse random seeds using the fine-tuned weights.}
\label{fig:compare_result}
\end{figure*}

\textbf{Dataset:} We evaluate the proposed approach on two tasks: defense against fine-tuning methods and defense against zero-shot methods. We employ the dataset constructed by Anti-DreamBooth, which consists of 50 individuals from CelebA-HQ~\cite{liu2015faceattributes} and 50 individuals from VGGFace2~\cite{cao2018vggface2}. Each individual is associated with 12 to 15 high-quality portraits of resolution $512\times512$. For experiments targeting fine-tuning method defense, we use the original resolution images, whereas in zero-shot defense experiments, the images are normalized to $112\times112$ via face alignment.

\noindent\textbf{Protected Model Selection:} We select the Stable Diffusion model as the base model. In fine-tuning defense experiments, we select SD-v2.1 as the surrogate model and further evaluate the transferability of protected images when applied to SD-v1.5. In zero-shot defense experiments, we select SD-v1.5 and SDXL to adapt diverse zero-shot adapters. 

\noindent\textbf{Selection of Customization Method:} During the optimization of perturbations $\delta_{ft}$, we use DreamBooth to fine-tune the surrogate models. We further evaluate the transferability of protected images when applied to LoRA. In zero-shot defense experiments, we select Faceid~\cite{ye2023ip-adapter} and Instance-ID~\cite{wang2024instantid} as the customization methods, which represent the two most widely adopted zero-sample identity imitation approaches. The inclusion of multiple customization methods enables an evaluation of the proposed approach's generalization capability.

\noindent\textbf{Hyperparameter Setting:} When optimizing $\delta_{ft}$ and $\delta_{zs}$, the perturbation bounds $\eta_{ft}$ and $\eta_{zs}$ are set to $7/255$ and $11/255$, respectively. The optimization strides $\sigma_{ft}$ and $\sigma_{zs}$ are $5 \times 10^{-3}$ and $8 \times 10^{-4}$. After generating protected images, we evaluate their protection performance by applying DreamBooth on these protected images to fine-tune pretrained weights. The fine-tuning is conducted with a batch size of 4, a total of 400 iterations, and a learning rate of $5 \times 10^{-6}$, a configuration that achieves strong identity similarity on unprotected images. 

\noindent\textbf{Comparison Methods:} For comparison, we select four state-of-the-art defense methods against diffusion model customization: MIST~\cite{liang2023adversarial}, Anti-DreamBooth~\cite{van2023anti} (denoted as Anti-DB), DisDiff~\cite{liu2024disrupting}, and Anti-diffusion~\cite{zheng2025anti}. For a fair comparison, the same hyperparameter settings are applied across all methods.

\subsection{Comparison Results of Fine-tuning Defense}
We evaluate the effectiveness of each defense method against fine-tuning of diffusion models based on two criteria: generated image quality and identity preservation. The comparative results across these methods are presented in Table~\ref{tab:compare}. For generated image quality, we employ the Fréchet Inception Distance~\cite{heusel2017gans} (FID), Efficient-FIQA~\cite{sun2025efficient} (denoted as FIQA), and Mean Opinion Score (MOS) from subjective assessment experiments as evaluation metrics. The details of subjective experiments are presented in supplementary materials. When the inference prompt (``\textit{a photo of sks person}") is identical to that used in optimizing perturbations, our approach significantly outperforms other methods in terms of FID and MOS on both datasets. When the inference prompt differs from the one used during optimization, our method still achieves superior performance on the CelebA-HQ dataset, but exhibits a slight decline on the VGGFace2 dataset. All methods achieve comparable scores on FIQA, with no significant differences observed. This may be attributed to the composition of the model's training dataset, which consists of naturally distorted images and high-quality AI-generated faces, rendering it less sensitive to the distortions introduced by these defense approaches.

In terms of identity preservation, we use Face Detection Rate (FDR) of RetinaFace detector~\cite{deng2020retinaface} and Identity Score Matching (ISM)~\cite{deng2019arcface} as evaluation metrics. The lower the values of these two metrics, the greater the deviation of the generated image from a human face, and the more distant the identity becomes relative to that in the fine-tuning dataset. Similar to generated image quality, our approach significantly outperforms other methods in terms of FID and ISM on both datasets when generating images using the same prompt as used during optimization. We present some visualization results in Figure~\ref{fig:compare_result} which demonstrates that our method places greater emphasis on protecting the facial region, leading to more pronounced degradation of facial details in the generated images.

\begin{table}
\caption{Robustness evaluation under transfer to LoRA. The best result is marked in \textbf{bold}.}
\centering
\begin{adjustbox}{width=0.45\textwidth}
\begin{tblr}{
  cells = {c},
  vline{2} = {-}{},
  hline{1,8} = {-}{0.08em},
  hline{2} = {-}{},
}
Method         & FDR$\downarrow$ & ISM$\downarrow$ & FID$\uparrow$ & FIQA$\downarrow$ & MOS$\downarrow$ \\
w/o Protect    & 0.942 & 0.433 & 58.51 & 0.386 & N/A  \\
MIST           & 0.988 & 0.388 & 86.11 & 0.295 & 3.48 \\
Anti-DB        & 0.776 & 0.280 & 152.3 & \textbf{0.240} & 2.39 \\
DisDiff        & 0.829 & 0.306 & 135.6 & 0.241 & 2.58 \\
Anti-diffusion & 0.825 & 0.298 & 146.6 & 0.265 & 2.72 \\
Ours           & \textbf{0.728} & \textbf{0.199} & \textbf{182.4} & 0.280 & \textbf{1.37}
\end{tblr}
\end{adjustbox}
\label{tab:lora}
\end{table}

\subsection{Robustness Results of Fine-tuning Defense}
Following prior work, we evaluate the robustness of defense methods along two dimensions: robustness to fine-tuning method variations and robustness to model version variations. In the first experiment, we fine-tune the dynamic surrogate model with DreamBooth during optimizing perturbations, and use LoRA for fine-tuning when testing the defense capability of the protected images. Table~\ref{tab:lora} demonstrates that our approach achieves the highest transferability to different fine-tuning methods compared to other methods. In the second experiment, we fine-tune the dynamic surrogate model based on SD-v2.1 during optimizing perturbations, and fine-tune a pretrained SD-v1.5 model when testing the protection performance. Table~\ref{tab:v1.5} demonstrates that our approach continues to significantly outperform the other methods when transferring to a different model version. The NSFWR in Table~\ref{tab:v1.5} denotes the detection rate of Not Safe For Work content in the generated results.


\begin{table}
\caption{Robustness evaluation under transfer to SD-v1.5.}
\centering
\begin{adjustbox}{width=0.41\textwidth}
\begin{tblr}{
  cells = {c},
  vline{2} = {-}{},
  hline{1,7} = {-}{0.08em},
  hline{2} = {-}{},
}
Method         & FDR$\downarrow$ & ISM$\downarrow$ & FID$\uparrow$ & NSFWR$\uparrow$ \\
MIST           & 0.604 & 0.313 & 213.3 & 0.263       \\
Anti-DB        & 0.606 & 0.323 & 239.4 & 0.341       \\
DisDiff        & 0.148 & 0.076 & 327.6 & 0.481       \\
Anti-diffusion & 0.258 & 0.139 & 312.1 & 0.493       \\
Ours           & \textbf{0.070} & \textbf{0.031} & \textbf{407.8} & \textbf{0.733}         
\end{tblr}
\end{adjustbox}
\label{tab:v1.5}
\end{table}

\begin{table}
\caption{Comparison results with state-of-the-art methods against zero-shot image-to-image generation.}
\centering
\begin{adjustbox}{width=0.46\textwidth}
\begin{tblr}{
  cells = {c},
  cell{1}{1} = {r=2}{},
  cell{1}{2} = {r=2}{},
  cell{1}{3} = {c=2}{},
  cell{1}{5} = {c=2}{},
  cell{3}{1} = {r=5}{},
  cell{8}{1} = {r=5}{},
  vline{3} = {1-12}{},
  vline{5} = {1-12}{},
  hline{1,13} = {-}{0.08em},
  hline{3,8} = {-}{},
}
Dataset   & Method         & Faceid &        & Instance-ID &        \\
          &                & ISM$_{pro}\downarrow$ & ISM$_{gen}\downarrow$ & ISM$_{pro}\downarrow$      & ISM$_{gen}\downarrow$ \\
CelebA-HQ & MIST           & 0.970 & 0.409 & 0.962 & 0.615 \\
          & Anti-DB        & 0.965 & 0.409 & 0.955 & 0.612 \\
          & DisDiff        & 0.951 & 0.405 & 0.960 & 0.606 \\
          & Anti-Diffusion & 0.971 & 0.412 & 0.968 & 0.606 \\
          & Ours           & \textbf{0.090} & \textbf{0.039} & \textbf{0.091} & \textbf{0.049}      \\
VGGFace2  & MIST           & 0.963 & 0.380 & 0.959 & 0.622 \\
          & Anti-DB        & 0.966 & 0.379 & 0.963 & 0.621 \\
          & DisDiff        & 0.965 & 0.375 & 0.960 & 0.620 \\
          & Anti-Diffusion & 0.968 & 0.381 & 0.965 & 0.620 \\
          & Ours           & \textbf{0.074} & \textbf{0.038} & \textbf{0.077} & \textbf{0.058}    
\end{tblr}
\end{adjustbox}
\label{tab:zero-shot}
\end{table}

\subsection{Comparison Results of Zero-shot Defense}
To evaluate the effectiveness of the proposed method in defending against zero-shot generation methods, we select the two most representative and widely used methods in the field of zero-shot facial identity synthesis as the target models: IP-Adapter Faceid (denoted as Faceid) based on SD-v1.5 and Instance-ID based on SDXL. The identity encoders of these two models utilize distinct pre-trained face recognition weights. We denote the identity similarity between the protected image and the original image as ISM$_{pro}$, and the identity similarity between the generated image and the original image as ISM$_{gen}$, respectively. As shown in Table~\ref{tab:zero-shot}, Existing methods exhibit high vulnerability to zero-shot generation techniques and provide minimal defensive capability. In contrast, the second perturbation layer in our dual-layer framework effectively prevents facial identity theft by zero-shot methods.

\begin{figure}[htbp]
\centering
\includegraphics[width=0.45\textwidth]{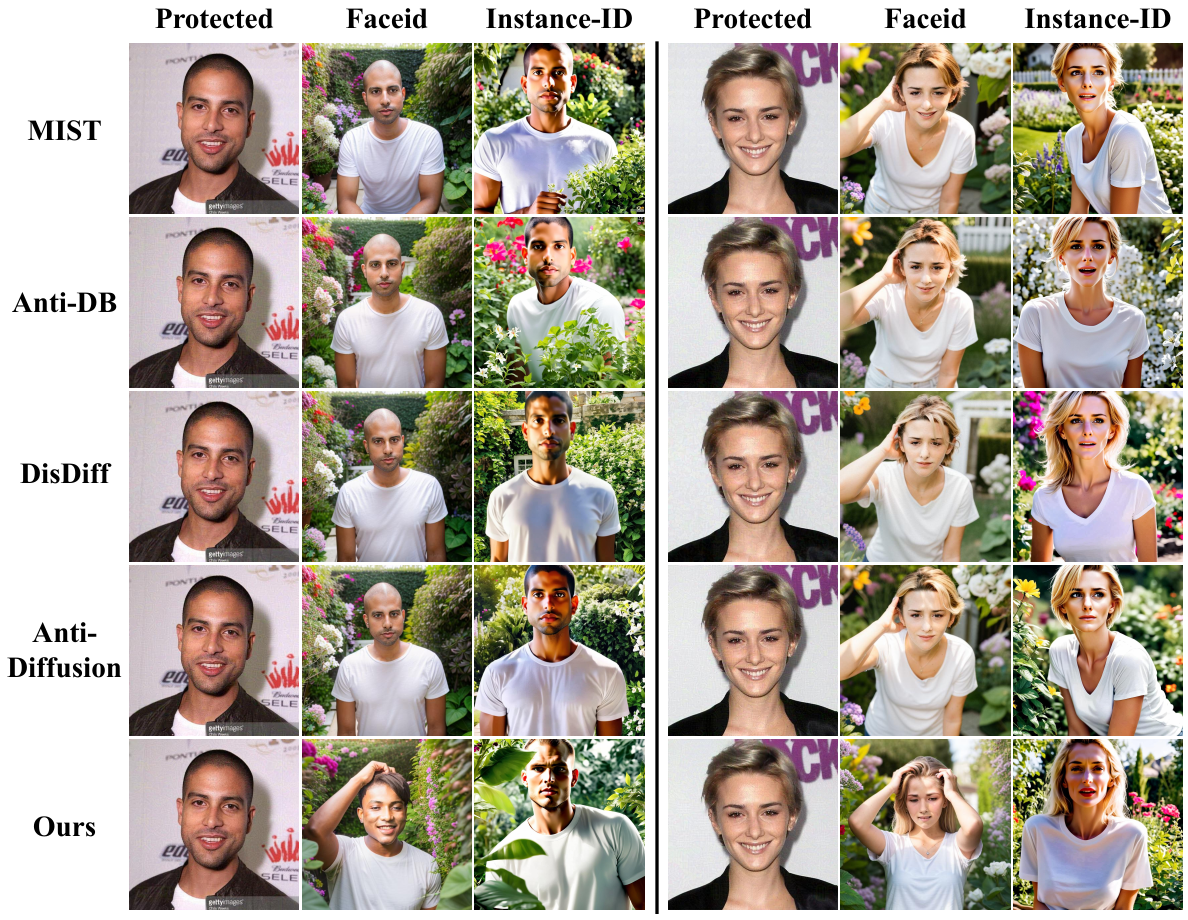}
\caption{The comparison results on defending Zero-shot generation methods.}
\label{fig:zero_shot_result}
\end{figure}


\subsection{Ablation Results}
Finally, we conduct ablation study to evaluate the proposed modules: Dual-Surrogate Models Mechanism (denoted as DSUR), Alternating Dynamic Fine-Tuning (denoted as ADFT), and perturbations for zero-shot methods (denoted as Anti-ZS). The contributions of each modules are presented in Table~\ref{tab:ablation}. As shown in Table~\ref{tab:compare}, the individual protective effect of each module surpasses that of most comparison methods.

\begin{table}
\caption{Results of ablation study.}
\centering
\begin{adjustbox}{width=0.45\textwidth}
\begin{tblr}{
  cells = {c},
  cell{1}{1} = {c=5}{},
  cell{6}{1} = {c=5}{}, 
  cell{7}{2} = {c=2}{},
  cell{7}{4} = {c=2}{},
  vline{2} = {2-5,7-10}{},
  vline{4} = {7-10}{},
  hline{1,11} = {-}{0.08em},
  hline{2,3,6,7,8,9} = {-}{},
}
Ablation Study for Anti-fine-tuning &        &        &        &        \\
Config                              & ISM$\downarrow$ & FDR$\downarrow$ & FID$\uparrow$ & FIQA$\downarrow$ \\
w/o DSUR                            & 0.160 & 0.316 & 213.7 & 0.236       \\
w/o ADFT                            & 0.277 & 0.607 & 180.9 & 0.244       \\
DSUR+ADFT                           & 0.096 & 0.201 & 233.7 & 0.221       \\
Ablation Study for Anti-zero-shot   &        &      &        &        \\
   & Faceid &      & Instance-ID &        \\
Config                              & ISM$_{pro}\downarrow$ & ISM$_{gen}\downarrow$ & ISM$_{pro}\downarrow$ & ISM$_{gen}\downarrow$ \\
w/o Anti-ZS                         & 0.974 & 0.398 & 0.965 & 0.618     \\
Anti-ZS                             & 0.082 & 0.039 & 0.054 & 0.084       
\end{tblr}
\end{adjustbox}
\label{tab:ablation}
\end{table}

\section{Conclusion}
\label{sec:conclusion}

This paper presents \textbf{D}ual-\textbf{L}ayer \textbf{A}nti-\textbf{Diff}usion (DLADiff), a dual-layer framework that defends against both fine-tuning and zero-shot customizations of diffusion models. The first layer prevents unauthorized fine-tuning using the Dual-Surrogate Models (DSUR) mechanism and Alternating Dynamic Fine-Tuning (ADFT). The second layer, though simple, effectively blocks zero-shot generation. Experiments show DLADiff outperforms existing methods in both defense scenarios.

\section{Appendix}

\begin{table}
\caption{Notion table of Models}
\centering
\begin{tblr}{
  cells = {c},
  vline{2} = {1-5}{},
  hline{1,6} = {-}{0.08em},
  hline{2} = {-}{},
}
Notion   & Definition                          \\
$\mathbf{Enc_{vae}}$ & encoder of VAE                      \\
$\mathbf{UNet_s}$  & UNet of the static surrogate model  \\
$\mathbf{UNet_s}$  & UNet of the dynamic surrogate model \\
$\mathbf{IE}$       & identity encoder     
\end{tblr}
\label{notion:model}
\end{table}

\begin{table}
\caption{Notion table of Operations}
\centering
\begin{tblr}{
  cells = {c},
  vline{2} = {1-11}{},
  hline{1,12} = {-}{0.08em},
  hline{2} = {-}{},
}
Notion   & Definition                          \\
$\mathbf{BP}$ & back propagation of gradients             \\
$\mathbf{Adam}$  & Adam optimizer  \\
$\mathrm{randint}$  & random integer generator \\
$\mathrm{randn}$       & normal noise generator  \\
$\mathrm{noise\_scheduler}$       & noise scheduler for forward diffusion  \\
$\nabla$       & gradients of relevant weights  \\
$\mathrm{affine\_transform}$       & affine transformation  \\
$\mathrm{face\_alignment}$       & face alignment  \\
$\mathrm{clip}$       & clip operation  \\
$\mathbf{CosSim}$       & cosine similarity  \\
\end{tblr}
\label{notion:operation}
\end{table}

\begin{table}
\caption{Notion table of Variables}
\centering
\begin{tblr}{
  cells = {c},
  vline{2} = {1-10}{},
  hline{1,10} = {-}{0.08em},
  hline{2} = {-}{},
}
Notion   & Definition                          \\
$x/x_{f}$ & image and face region of the image \\
$z$  & latent variables  \\
$\epsilon$ & random noises \\
$\theta_{pre}/\theta_d/\theta_s$       & pretrained/dynamic/static weights  \\
$\delta_{ft}/\delta_{zs}$       & the first and second perturbations  \\
$\eta_{ft}/\eta_{zs}$       & the perturbation bounds   \\
$\mathrm{M_{affine}}$       & affine matrix \\
$ths$ & threshold value of similarity
\end{tblr}
\label{notion:variable}
\end{table}

\begin{algorithm}[!ht]
    \caption{Perturbation Optimization for Fine-tuning}
    \label{alg:ft}
    \renewcommand{\algorithmicrequire}{\textbf{Input:}}
    \renewcommand{\algorithmicensure}{\textbf{Output:}}
    \begin{algorithmic}[1]
        \REQUIRE $\mathcal{X},\mathcal{X}_{clean}$  
        \ENSURE $\delta_{ft}$    
        \STATE  $\theta_s=\theta_{pre},\ \theta_d=\theta_{pre},\ \delta_{ft}=\mathbf{0}$
        \WHILE{$iter<iter_{ft}$}
            \STATE  $z_{0}=\mathbf{Enc_{vae}}(x),\ s.t.\ x\in\mathcal{X}_{clean}$
            \STATE  $t=\mathrm{randint}(1,999),\ \epsilon=\mathrm{randn}(z_0.\mathrm{shape})$
            \STATE  $z_{t+1}=\mathrm{noise\_scheduler}(z_0,t,\epsilon)$
            \STATE  $\mathcal{L}_{cond}=\mathbb{E}_{z_0,t,c}||\epsilon-\mathbf{UNet_s}(z_{t+1},t,c,\theta_s)||^2_2$
            \STATE $\nabla_{\theta_s}=\mathbf{BP}(\mathcal{L}_{cond},\theta_s)$
            \STATE $\theta_s=\mathbf{Adam}(\theta_s,\nabla_{\theta_s})$
        \ENDWHILE
        \WHILE{$iter<iter_{opt}$}
            \FOR{$i \ \in [1, iter_1]$}
                \STATE  $z_{0}=\mathbf{Enc_{vae}}(x),\ s.t.\ x\in\mathcal{X}$
                \STATE
                $z'_{0}=\mathbf{Enc_{vae}}(x+\delta_{ft}),\ s.t.\ x\in\mathcal{X}$
                \STATE  $t=\mathrm{randint}(1,999),\ \epsilon=\mathrm{randn}(z_0.\mathrm{shape})$
                \STATE  $z_{t+1}=\mathrm{noise\_scheduler}(z_0,t,\epsilon)$
                \STATE  $z'_{t+1}=\mathrm{noise\_scheduler}(z'_0,t,\epsilon)$
                \STATE  $\mathcal{M}c_{\theta_s}(x)=\mathbf{UNet_s}(z_{t+1},t,c,\theta_s).\mathrm{att_1}$
                \STATE  $\mathcal{M}c_{\theta_d}(x+\delta_{ft})=\mathbf{UNet_d}(z'_{t+1},t,c,\theta_d).\mathrm{att_1}$
                \STATE  $\mathcal{M}s_{\theta_s}(x)=\mathbf{UNet_s}(z_{t+1},t,c,\theta_s).\mathrm{att_2}$
                \STATE  $\mathcal{M}s_{\theta_d}(x+\delta_{ft})=\mathbf{UNet_d}(z'_{t+1},t,c,\theta_d).\mathrm{att_2}$
                \STATE $\mathcal{L}_{att}=||\mathcal{M}c_{\theta_s}(x)-\mathcal{M}c_{\theta_d}(x+\delta_{ft})||^2_2
+||\mathcal{M}s_{\theta_s}(x)-\mathcal{M}s_{\theta_d}(x+\delta_{ft})||^2_2,$
                \STATE $\nabla_{\delta_{ft}}=\mathbf{BP}(\mathcal{L}_{att},\delta_{ft})$
                \STATE $\delta_{ft}=\mathrm{clip}(\delta_{ft}+\sigma_{ft}*\nabla_{\delta_{ft}},\ -\eta_{ft},\ +\eta_{ft})$,
            \ENDFOR
            \FOR{$i \ \in [1, iter_2]$}
                \STATE  $ z_{0}=\mathbf{Enc_{vae}}(x+\delta_{ft}),\ s.t.\ x\in\mathcal{X}$
                \STATE  $t=\mathrm{randint}(1,999),\ \epsilon=\mathrm{randn}(z_0.\mathrm{shape})$
                \STATE  $z_{t+1}=\mathrm{noise\_scheduler}(z_0,t,\epsilon)$
                \STATE $\mathcal{L}_{cond}=\mathbb{E}_{z_0,t,c}||\epsilon-\mathbf{UNet_d}(z_{t+1},t,c,\theta_d)||^2_2$
                \STATE $\nabla_{\delta_{ft}}=\mathbf{BP}(\mathcal{L}_{cond},\delta_{ft})$
                \STATE $\delta_{ft}=\mathrm{clip}(\delta_{ft}+\sigma_{ft}*\nabla_{\delta_{ft}},\ -\eta_{ft},\ +\eta_{ft})$
            \ENDFOR
            \FOR{$i \ \in [1, iter_3]$}
                \STATE  $ z_{0}=\mathbf{Enc_{vae}}(x+\delta_{ft}),\ s.t.\ x\in\mathcal{X}$
                \STATE  $t=\mathrm{randint}(1,999),\ \epsilon=\mathrm{randn}(z_0.\mathrm{shape})$
                \STATE  $z_{t+1}=\mathrm{noise\_scheduler}(z_0,t,\epsilon)$
                \STATE $\mathcal{L}_{cond}=\mathbb{E}_{z_0,t,c}||\epsilon-\mathbf{UNet_d}(z_{t+1},t,c,\theta_d)||^2_2$
                \STATE $\nabla_{\theta_d}=\mathbf{BP}(\mathcal{L}_{cond},\theta_d)$
                \STATE $\theta_d=\mathbf{Adam}(\theta_d,\nabla_{\theta_d})$
            \ENDFOR
        \ENDWHILE
        \RETURN \textbf{Output}
    \end{algorithmic}
\end{algorithm}

\begin{algorithm}[!ht]
    \caption{Perturbation Optimization for Zero-shot}
    \label{alg:zs}
    \renewcommand{\algorithmicrequire}{\textbf{Input:}}
    \renewcommand{\algorithmicensure}{\textbf{Output:}}
    \begin{algorithmic}[1]
        \REQUIRE $\delta_{ft},\ \mathcal{X}$  
        \ENSURE $\delta_{zs}$    
        \STATE  $\delta_{zs}=\mathbf{0}$
        \STATE  $x'=x+\delta_{ft},\ s.t.\ x\in\mathcal{X}$
        \WHILE{$\mathbf{CosSim}(\mathcal{E}_{pro\_i}, \mathcal{E}_{tar\_i})>ths$}
            \STATE $x_f,\ \mathrm{M_{affine}}=\mathrm{face\_alignment}(x)$
            \STATE $\epsilon=\mathrm{randn}(\mathrm{M_{affine}.shape})$
            \STATE $\mathrm{M'_{affine}}=\mathrm{M_{affine}}+\epsilon$
            \STATE $x'_f=\mathrm{affine\_transform}(x',\ \mathrm{M'_{affine}})$
            \STATE $\mathcal{L}_{id}=1$-$\sum^{N}_{i=1} \mathbf{CosSim}(\mathbf{IE}_{i}(x'_f+\delta_{zs}), \mathbf{IE}_{i}(x_f))$
            \STATE  $\delta_{zs}=\delta_{zs}+\sigma_{zs}*\nabla_{\mathcal{\delta}_{zs}}\mathcal{L}_{id}$
            \STATE  $\delta_{zs}=\mathrm{clip}(\delta_{zs}, \mathrm{min}$=$-\eta_{zs}, \mathrm{max}$=$+\eta_{zs})$
        \ENDWHILE
        \RETURN Outputs
    \end{algorithmic}
\end{algorithm}

\subsection{Algorithm Details}
We present the detailed pipelines of the two-layer perturbation optimization in Algorithm~\ref{alg:ft} and Algorithm~\ref{alg:zs}. The notions used in these algorithms are defined in Table~\ref{notion:model}, Table~\ref{notion:operation}, and Table~\ref{notion:variable}. Table~\ref{notion:model} defines the models, Table~\ref{notion:operation} defines the operations, and Table~\ref{notion:variable} defines the variables used in these two algorithms.

\subsection{Subjective Assessment Experiments}
\begin{figure}[htbp]
\centering
\includegraphics[width=0.45\textwidth]{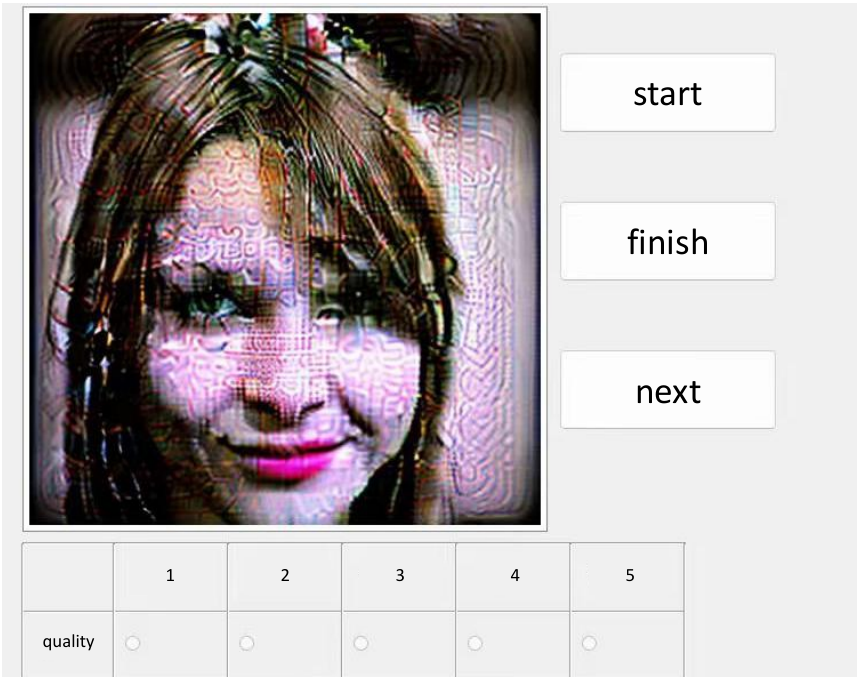}
\caption{The user interface of subjective assessment experiments.}
\label{fig:mos_ui}
\end{figure}

\begin{figure*}[htbp]
\centering
\includegraphics[width=0.9\textwidth]{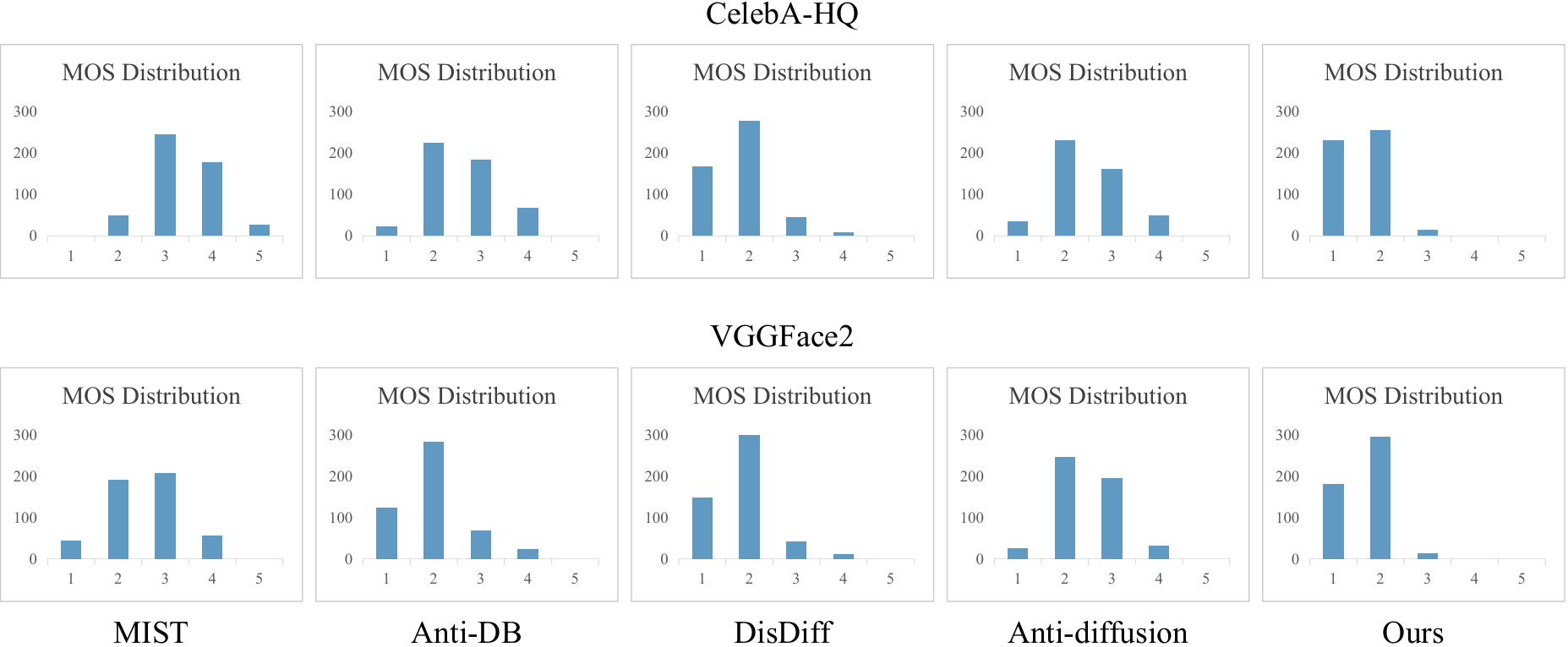}
\caption{The distributions of Mean Opinion Score (MOS).}
\label{fig:mos_score}
\end{figure*}

To obtain Mean Opinion Score (MOS), we invite ten volunteers including five male and five female participants to assess the visual quality of the images generated after fine-tuning on the protected images with each defense methods. Visual quality is defined on a five-point scale ranging from low to high, where a score of 5 corresponds to the generation results obtained using fine-tuning with clean images, and scores from 4 to 1 represent progressively increasing levels of distortion. The detailed scoring criteria are as follows:
\begin{itemize}
\item ``4'': The image has slight distortions, such as artifacts, blurring, noise, and blocky distortions.
\item ``3'': The distortions are more severe than ``4'' but the facial features such as eyes and mouse can still be recognized.
\item ``2'': The facial features and details are significantly destroyed.
\item ``1'': Unrecognizable or disgusting, terrifying faces.
\end{itemize}
We randomly select 50 faces generated from the fine-tuned models that are fine-tuned on the protected images from each defense methods. The user interface (UI) of this experiment is presented in Figure~\ref{fig:mos_ui}. We present the distributions of MOS in Figure~\ref{fig:mos_score}.

\subsection{Supplementary Explanation of Experiments}
As stated in the main text, the same perturbation bound for fine-tuning is applied to all comparison methods to ensure a fair and consistent evaluation. However, our approach incorporates two layers of perturbation, whereas other methods employ only a single layer designed to defend against fine-tuning methods. Therefore, we report the peak signal to noise ratio (PSNR) of the protected images generated by each comparison method. As shown in Table~\ref{tab:psnr}, all methods achieve a comparable PSNR ($35\pm0.5$). 

\begin{table}
\caption{PSNR of the protected images generated by comparison methods.}
\centering
\begin{adjustbox}{width=0.45\textwidth}
\begin{tabular}{c|ccccc} 
\toprule
     & MIST  & Anti-DB & DisDiff & Anti-diffusion & Ours   \\ 
\hline
PSNR & 34.61 & 34.52   & 35.16   & 35.24          & 35.06  \\
\bottomrule
\end{tabular}
\end{adjustbox}
\label{tab:psnr}
\end{table}

In Table 1 of the main text, the FID scores are computed between two image datasets: (1) the clean image dataset to be protected, (2) the generated images using the weights fine-tuned on protected images. For each individual in CeleA-HQ or VGGFace2, the images to be protected contain four samples. Therefore, the first dataset includes 200 images in total. For the second dataset, we generate 20 images for each individual using diverse random seeds. Therefore, the second dataset includes 1000 images. In contrast to our experiments, the FID scores reported in other papers~\cite{liu2024disrupting} are computed for each individual. Since the Fréchet Inception Distance (FID) measures the distance between two probability distributions, a higher number of samples in both datasets leads to a more accurate estimation with reduced statistical error. Consequently, our evaluation approach yields more reliable results and consistently reports lower FID scores compared to those presented in prior studies. We also present the results computed according to~\cite{liu2024disrupting} in Table~\ref{tab:fidbig}. The DreamBooth results on CelebA-HQ and VGGFace2 are denoted as DB-C and DB-V, respectively.

\begin{table}
\caption{FID results using small-scale datasets}
\centering
\begin{adjustbox}{width=0.45\textwidth}
\begin{tblr}{
  cells = {c},
  vline{2} = {-}{},
  hline{1,5} = {-}{0.08em},
  hline{2} = {-}{},
}
     & MIST & Anti-DB & DisDiff & Anti-diffusion & Ours \\
DB-C & 168.0 & 220.1 & 286.2 & 239.7 & \textbf{320.7} \\
DB-V & 306.8 & 306.6 & 293.4 & 298.0 & \textbf{314.4} \\
LoRA & 172.3 & 243.7 & 230.3 & 232.5 & \textbf{284.1}
\end{tblr}
\end{adjustbox}
\label{tab:fidbig}
\end{table}

\subsection{More Visualization Results}
\begin{figure}[htbp]
\centering
\includegraphics[width=0.42\textwidth]{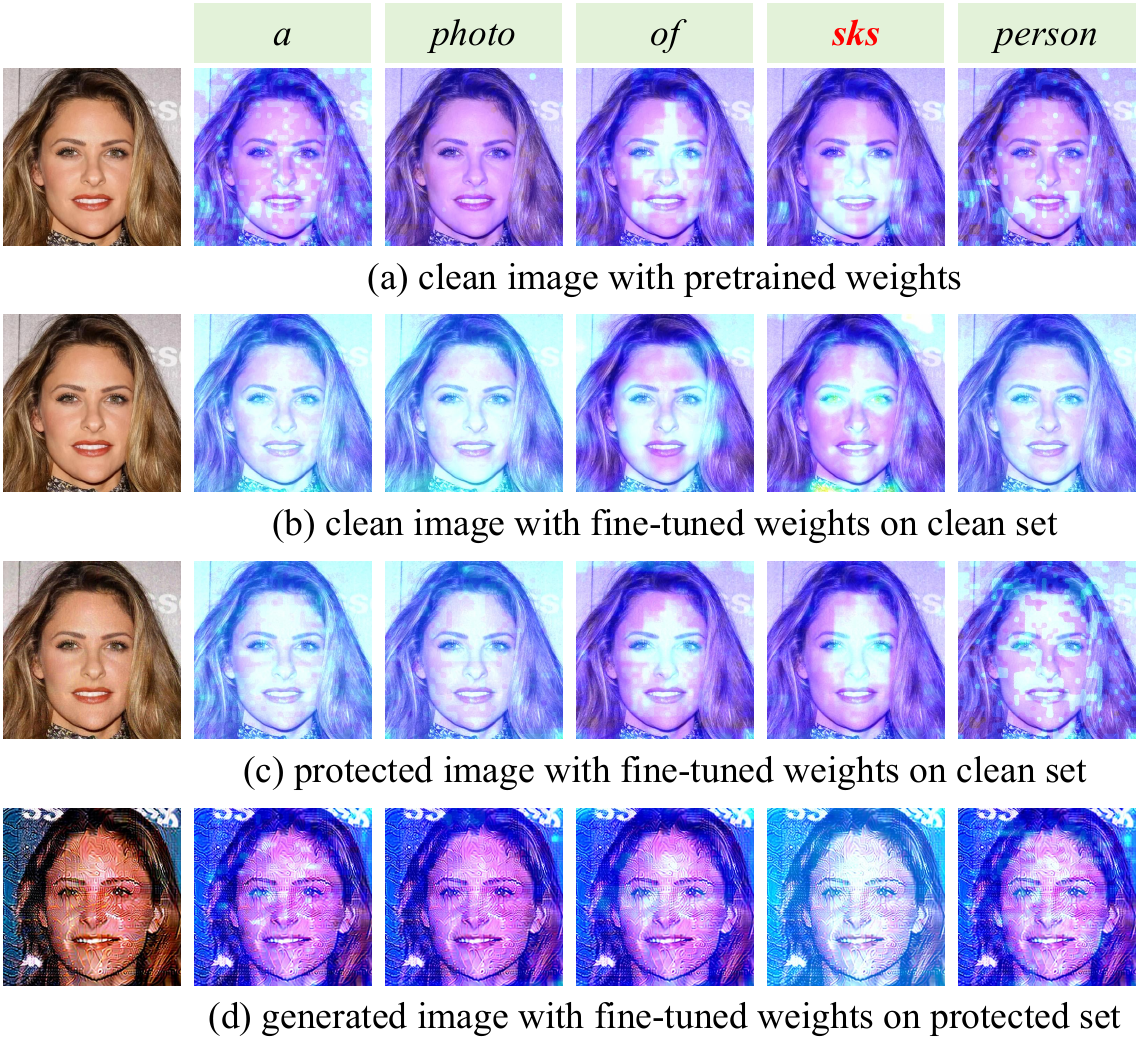}
\caption{The visualization results of cross-attention maps.}
\label{fig:atten_map}
\end{figure}

In this section, we present more visualization results. We visualize the cross-attention maps to demonstrate the effectiveness of our approach to defend fine-tuning. As illustrated in Figure~\ref{fig:atten_map}~(a), the cross-attention maps of the unprotected clean image, when is processed using pre-trained weights without fine-tuning, exhibits no focused attention on specific regions. In contrast, as shown in Figure~\ref{fig:atten_map}~(b), the same clean image processed with weights fine-tuned on clean images reveals a clear correspondence between special tokens (``\textit{sks}'') and specific image regions (eyes, nose, and mouse). In Figure~\ref{fig:atten_map}~(c), we use the weights fine-tuned on clean images to process the protected image. The cross-attention maps exhibit significant differences compared to the results shown in Figure~\ref{fig:atten_map}~(b), indicating that the perturbations effectively disrupts the normal cross-attention mechanism. Figure~\ref{fig:atten_map}~(d) further illustrates that there is no focused attention on specific regions in generated images.

\begin{figure}[htbp]
\centering
\includegraphics[width=0.42\textwidth]{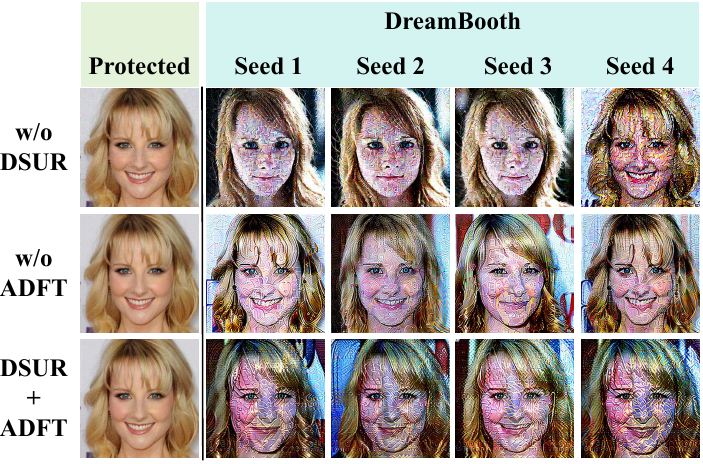}
\caption{The ablation results of the first layer.}
\label{fig:ablation}
\end{figure}

Figure~\ref{fig:ablation} presents the visualization results of ablation study for the first protective layer. As shown in Figure~\ref{fig:ablation}, DSUR focuses more on introducing high-frequency textures to disrupt local facial details, while ADFT is more significant in degrading the overall quality.


\begin{figure*}[htbp]
\centering
\includegraphics[width=0.95\textwidth]{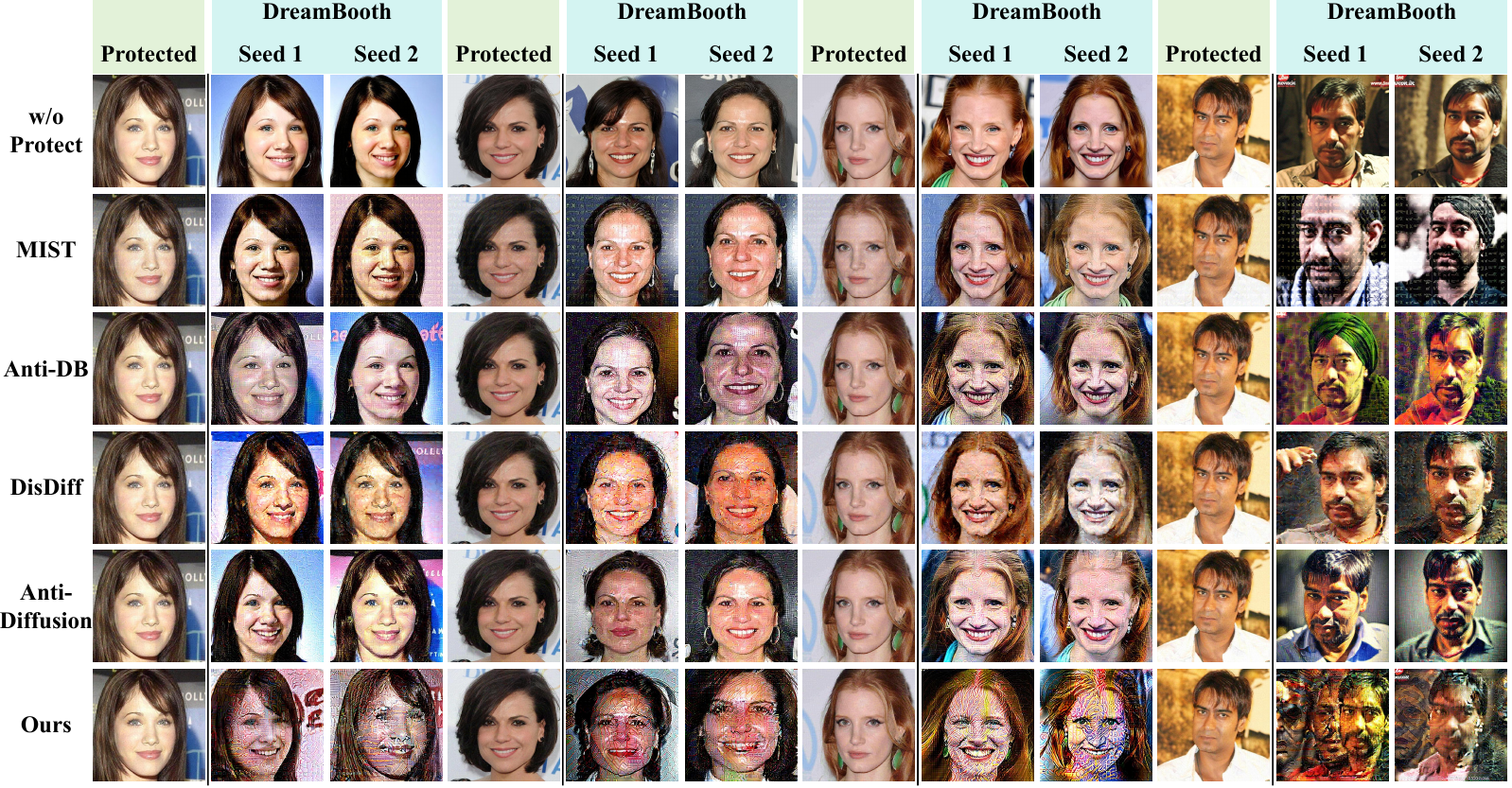}
\caption{The comparison results on defending DreamBooth fine-tuning. We use the protected images to fine-tune a pretrained
stale diffusion model. Then, we generate images under diverse random seeds using the fine-tuned weights.}
\label{fig:compare-db}
\end{figure*}

\begin{figure*}[htbp]
\centering
\includegraphics[width=0.95\textwidth]{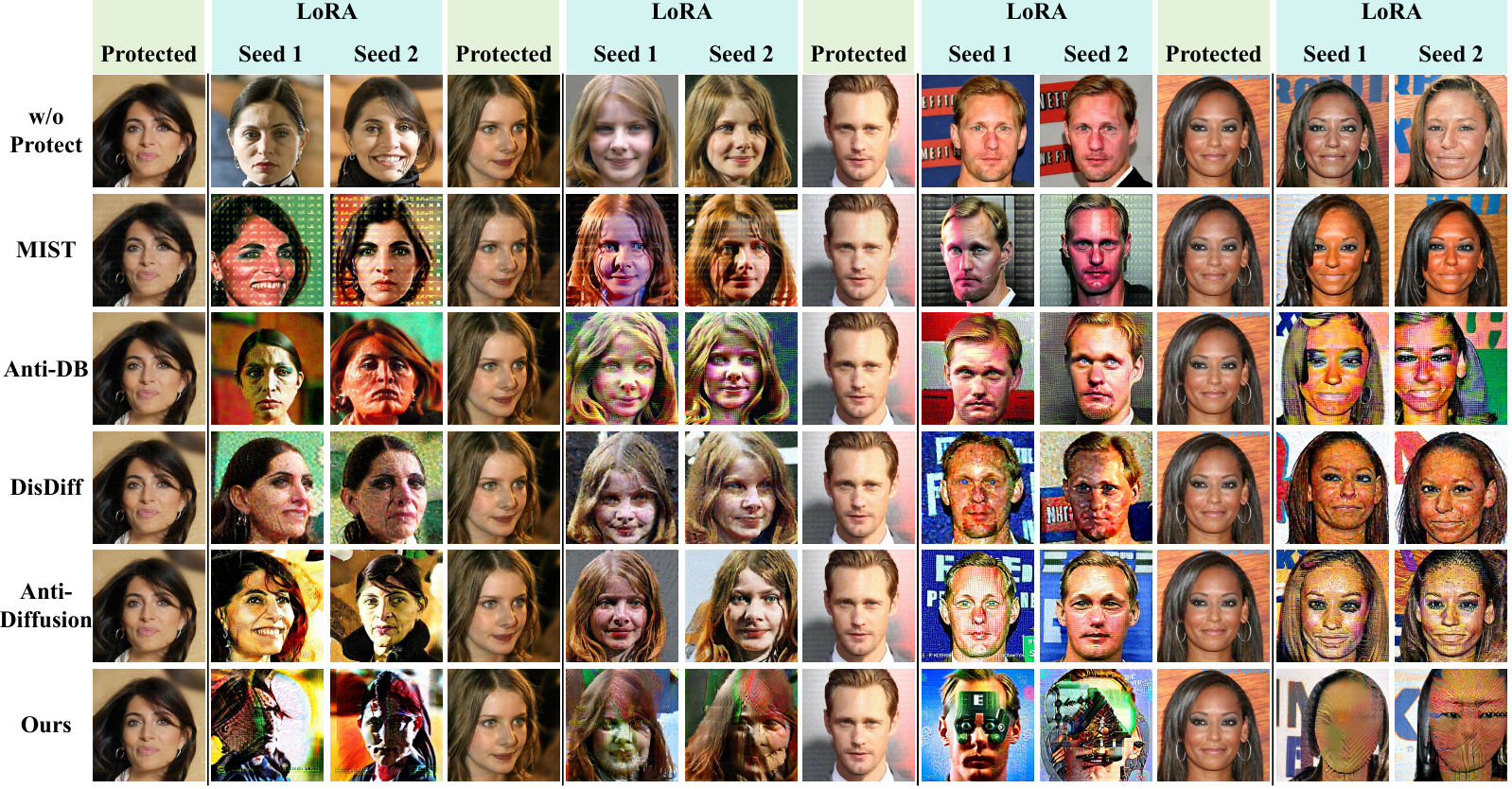}
\caption{The comparison results on defending LoRA fine-tuning. We use the protected images to fine-tune a pretrained
stale diffusion model. Then, we generate images under diverse random seeds using the fine-tuned weights.}
\label{fig:compare-lora}
\end{figure*}

\begin{figure*}[htbp]
\centering
\includegraphics[width=0.95\textwidth]{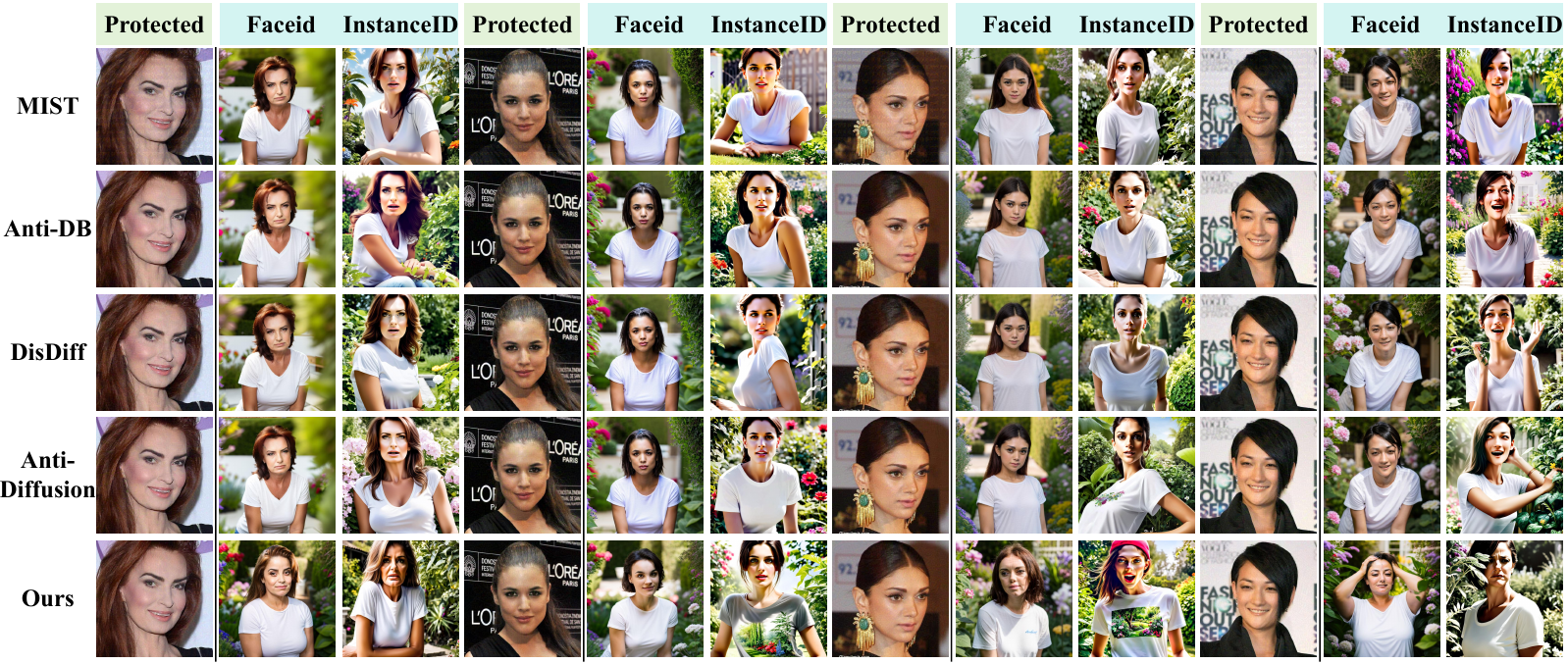}
\caption{The comparison results on defending zero-shot generation. Text prompts: ``\textit{a young woman in white T-shirt in a garden}'' and ``\textit{best quality, high quality, a wooden house in forest}''.}
\label{fig:compare-zero}
\end{figure*}

Figure~\ref{fig:compare-db} presents the visualization results of defending DreamBooth fine-tuning. Figure~\ref{fig:compare-lora} presents the visualization results of defending LoRA fine-tuning. Figure~\ref{fig:compare-zero} presents the visualization results of defending zero-shot generation.


{
    \small
    \bibliographystyle{ieeenat_fullname}
    \bibliography{main}
}

\end{document}